\documentclass[usenatbib,referee]{mn2e}
\usepackage{graphicx}
 
\begin{document}
  
\title{Photometric and spectroscopic evolution of the type IIP supernova 
SN 2004et}
 
\author[Sahu, D. K. et al.]{D. K. Sahu\thanks{E-mail : dks@crest.ernet.in (DKS)},
G. C. Anupama\thanks{E-mail : gca@iiap.res.in (GCA)}, S. Srividya, S. Muneer\\
Indian Institute of Astrophysics, Koramangala, Bangalore 560 034, India }

\date{Received / Accepted}
                                                                                
\pagerange{\pageref{firstpage}--\pageref{lastpage}} \pubyear{2006}
\maketitle

 
\begin{abstract}
We present optical photometry and spectroscopy of the type IIP supernova 
SN 2004et that occurred in the nearby galaxy NGC 6946. The observations span a 
time range of 8 days to 541 days after explosion. The late time bolometric 
luminosity and the H$\alpha$ luminosity in the nebular phase indicate that
$0.06\pm0.02{M_\odot}$ of $^{56}$Ni was synthesised during the 
explosion. The plateau luminosity, its duration and the expansion velocity of 
the supernova at the middle of the plateau indicate an explosion energy of
$E_{\rm{exp}}= 1.20^{+0.38}_{-0.30} \times 10^{51}$ ergs. The late time light 
curve and the evolution of the [OI]  and H$\alpha$ emission line profiles 
indicate the possibility of an early dust formation in the supernova ejecta.
 The luminosity of  [OI] 6300, 6364~\AA \, doublet, before the dust formation phase, is found 
to be comparable  to that of SN 1987A at similar epochs, impling an oxygen mass in the 
range $1.5 - 2{M_\odot}$, and a main sequence mass of $20{M_\odot}$ for the  progenitor.  

\end{abstract}
\begin{keywords}
supernovae: general - supernovae: individual: SN 2004et - galaxies:
individual: NGC 6946
\end{keywords}

\section{Introduction}
\label{intro}
Supernovae have been classified mainly based on their spectrum near maximum, 
the events which show the presence of hydrogen lines have been termed as 
supernovae type II (SNe II) and those which do not show hydrogen are classified
as type I. Type II events are thought to arise from the gravitational collapse 
of stars more massive than $8\,M_\odot$. SNe II have been further divided 
based on their light curve \citep{baronr} as type II-L (linear) and type II-P 
(plateau). Supernovae type IIP (SNe IIP) are characterized by a plateau of
nearly constant luminosity in their light curve, originating because of
propagation of a cooling and recombination wave through the supernova
envelope.

\cite{hamuy03} has shown that SNe IIP form a sequence from low-luminosity, 
low-velocity, nickel-poor events to bright, high velocity, nickel-rich objects. 
Further, there is an indication that more massive progenitors produce more 
energetic explosion and supernovae with greater energies produce more nickel. 
The direct identification of the progenitor of a few SNe IIP indicates these 
events arise from stars with masses in the range $\sim8\,-15\,{\rm{M}}_\odot$ 
(\citealt{smartt}, \citealt{van},  \citealt{li3}).

SN 2004et was discovered by Moretti on Sept. 27, 2004 in the nearby starburst 
galaxy NGC 6946, which has already produced seven supernovae during  1917 to 
2003. Based on a high resolution
spectrum that showed a relatively featureless spectrum with a very broad, low 
contrast H$\alpha$ emission, \cite{zwitter} classified the supernova as a 
type II event, which  was later confirmed by a low resolution spectrum taken on 
2004 Oct 01 by \cite{filippenko04}. The P-Cygni profile of H$\alpha$ was greatly 
dominated by the emission component, while the other hydrogen Balmer lines had
a more typical P-Cygni profile. The continuum was quite blue, although there was
very low flux shortward of 4000\AA. The supernova was detected in the radio 
frequencies at 22.460 GHz and 8.460 GHz on 2004 Oct 5.128 \citep{stockdale}, 
suggesting the presence of appreciable circumstellar material  around 
SN 2004et.  Based on pre-explosion images of NGC 6946, \cite{li1} identified 
the candidate progenitor as a yellow supergiant with an estimated zero age main 
sequence  mass of ${15}^{+5}_{-2}{M_\odot}$. The identification of the 
progenitor was confirmed with post-outburst Hubble Space Telescope images 
obtained on 2005 May 02 \citep{li2}. This makes SN 2004et one of the few 
core-collapse supernovae with a directly identified progenitor. 

The proximity and brightness of SN 2004et made it an ideal target for an
intensive monitoring. We present in this paper the results based on 
extensive photometric and spectroscopic observations of SN 2004et during $\sim$ 8 - 541 days
 since explosion.
 
\section{Observations and Data Reduction}
\subsection{Photometry}
SN 2004et was observed with the 2m Himalayan Chandra Telescope (HCT) of the
Indian Astronomical Observatory (IAO), Hanle, India, using the Himalaya Faint
Object Spectrograph Camera (HFOSC), equipped with a $2048\times 4096$ pixel CCD.
The central $2048\times 2048$ region of the CCD used for imaging covers a field 
of view of 10x10 arcmin, with a scale of 0.296 arcsec pixel$^{-1}$. The 
photometric monitoring of SN 2004et, in Bessell $UBVRI$ filters \citep{bessell1}, began on 2004
September 29, soon after its discovery, and continued until 2006 March 16.

\begin{figure}
\resizebox{\hsize}{!}{\includegraphics{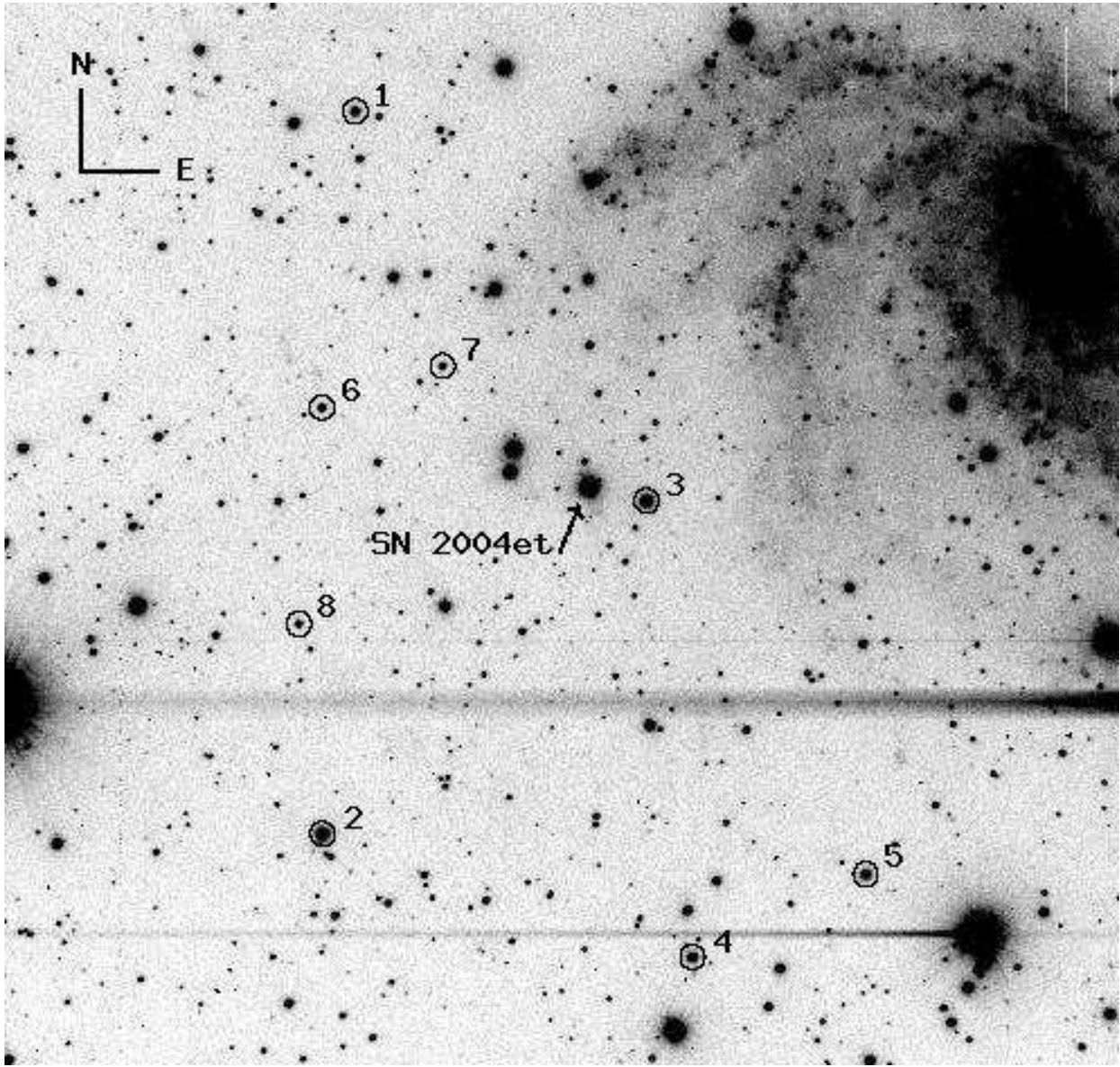}}
\caption[]{Identification chart for SN 2004et. The stars used as local standards are
marked as numbers 1-8. The field of view
is $10^\prime \times 10^\prime$.}
\label{fig1}
\end{figure}

The data were bias subtracted, flat-field corrected and cosmic rays removed
adopting the standard manner, using the various tasks available under the
IRAF software. Data obtained on photometric nights were calibrated using 
standard fields \citep{landolt} and a sequence of local standards in the 
supernova field (ref. Fig. \ref{fig1}) were used for photometric calibration of the
supernova. Table \ref{tab1} gives the $U, B, V, R, I$ magnitudes of the 
secondary standards averaged over a few photometric nights. 

Aperture photometry was performed on the local standards using an aperture of
radius determined based on an aperture growth curve. The magnitudes of the 
supernova and also the local standards were estimated using the profile-fitting 
method, using a fitting radius corresponding to the full width at half maximum
of the stellar profile. The difference between aperture and profile-fitting 
magnitudes was obtained using the standards and this correction was applied to 
the supernova magnitude. The supernova magnitudes were calibrated differentially
with respect to the local standards listed in Table \ref{tab1},  the  final magnitude
of the supernova was derived by taking an average of these estimates.  The estimated 
supernova magnitudes and errors are listed in Table \ref{tab2}. The errors in the  magnitudes
were estimated by combining the fit errors in quardature with those introduced by the 
transformation of instrumental magnitudes into the standard system. 

\begin{table*}
\begin{minipage}{126mm}
\caption{Magnitudes for the sequence of secondary standard stars in the field of SN 2004et.
The stars are identified in Fig. \ref{fig1}.}
\label{tab1}
\centering
\begin{tabular}{lccccc}
\hline\hline
ID & U  & B & V &  R & I \\
\hline
1&  $15.927\pm 0.020$ &  $15.828\pm 0.008$ &  $15.130\pm 0.009$ &  $14.672\pm 0.009$ & $14.219\pm 0.009$\\
2&  $14.439\pm 0.027$ &  $14.415\pm 0.010$ &  $13.773\pm 0.009$ &  $13.353\pm 0.007$ & $12.936\pm 0.011$\\
3&  $16.665\pm 0.021$ &  $15.521\pm 0.010$ &  $14.265\pm 0.008$ &  $13.568\pm 0.011$ & $12.903\pm 0.008$\\
4&  $15.840\pm 0.024$ &  $15.544\pm 0.017$ &  $14.726\pm 0.016$ &  $14.255\pm 0.016$ & $13.802\pm 0.009$\\
5&  $15.780\pm 0.021$ &  $15.582\pm 0.012$ &  $14.795\pm 0.011$ &  $14.336\pm 0.007$ & $13.854\pm 0.008$\\
6&  $17.485\pm 0.033$ &  $16.910\pm 0.012$ &  $15.953\pm 0.012$ &  $15.399\pm 0.012$ & $14.866\pm 0.009$\\
7&  $16.957\pm 0.028$ &  $16.827\pm 0.015$ &  $16.202\pm 0.011$ &  $15.782\pm 0.015$ & $15.365\pm 0.025$\\
8&  $16.865\pm 0.034$ &  $16.870\pm 0.017$ &  $16.233\pm 0.011$ &  $15.842\pm 0.015$ & $15.422\pm 0.015$\\

\hline
\end{tabular}
\end{minipage}
\end{table*}

\begin{table*}
\begin{minipage}{126mm}
\caption{Photometric observations of SN 2004et}
\label{tab2*}
\centering
\begin{tabular}{lccccccc}
\hline\hline
Date & J.D. & Phase\rlap{*} & U &  B & V &  R &  I\\
     & 2453000+ & (days) & \\
\hline
   29/09/2004&   278.202&  7.70 & $12.308\pm 0.035$&   $12.900\pm 0.018$&  $12.650\pm 0.013$&  $12.360\pm 0.021$&                    \\
   30/09/2004&   279.225&  8.73 & $12.224\pm 0.034$&   $12.932\pm 0.026$&  $12.633\pm 0.021$&  $12.331\pm 0.012$&                    \\
   01/10/2004&   280.200&  9.70 & $12.254\pm 0.034$&   $12.907\pm 0.017$&  $12.629\pm 0.016$&  $12.314\pm 0.013$&                    \\
   06/10/2004&   285.118& 14.62 & $12.418\pm 0.032$&   $12.908\pm 0.012$&  $12.556\pm 0.012$&  $12.216\pm 0.014$&  $ 12.019\pm 0.018$\\
   07/10/2004&   286.065& 15.57 & $12.514\pm 0.035$&   $12.921\pm 0.016$&  $12.547\pm 0.013$&  $12.200\pm 0.026$&  $ 12.009\pm 0.014$\\
   09/10/2004&   288.059& 17.56 & $12.563\pm 0.038$&   $12.963\pm 0.016$&  $12.549\pm 0.022$&  $12.182\pm 0.018$&  $ 11.962\pm 0.014$\\
   11/10/2004&   290.060& 19.56 & $12.673\pm 0.034$&   $12.998\pm 0.017$&  $12.552\pm 0.011$&  $12.156\pm 0.019$&  $ 11.937\pm 0.019$\\
   12/10/2004&   291.237& 20.74 & $12.751\pm 0.035$&   $13.047\pm 0.012$&  $12.560\pm 0.019$&  $12.167\pm 0.014$&  $ 11.927\pm 0.024$\\
   13/10/2004&   292.077& 21.58 & $12.793\pm 0.033$&   $13.065\pm 0.017$&  $12.558\pm 0.013$&  $12.161\pm 0.020$&  $ 11.923\pm 0.022$\\
   14/10/2004&   293.103& 22.60 & $12.885\pm 0.037$&   $13.113\pm 0.021$&  $12.560\pm 0.024$&  $12.160\pm 0.017$&  $ 11.918\pm 0.019$\\
   15/10/2004&   294.114& 23.61 & $12.997\pm 0.035$&   $13.162\pm 0.009$&  $12.557\pm 0.019$&  $12.154\pm 0.020$&  $ 11.922\pm 0.012$\\
   16/10/2004&   295.125& 24.62 & $13.063\pm 0.046$&   $13.203\pm 0.017$&  $12.568\pm 0.023$&  $12.156\pm 0.015$&  $ 11.912\pm 0.021$\\
   17/10/2004&   296.066& 25.57 & $13.178\pm 0.042$&   $13.240\pm 0.013$&  $12.572\pm 0.014$&  $12.157\pm 0.016$&  $ 11.892\pm 0.018$\\
   22/10/2004&   301.046& 30.55 & $13.638\pm 0.046$&   $13.501\pm 0.028$&  $12.656\pm 0.020$&  $12.210\pm 0.007$&  $ 11.916\pm 0.016$\\
   23/10/2004&   302.045& 31.55 & $13.737\pm 0.043$&   $13.573\pm 0.021$&  $12.646\pm 0.023$&  $12.207\pm 0.021$&  $ 11.922\pm 0.013$\\
   26/10/2004&   305.083& 34.58 & $13.865\pm 0.048$&   $13.694\pm 0.026$&  $12.704\pm 0.016$&  $12.249\pm 0.020$&  $ 11.936\pm 0.019$\\
   27/10/2004&   306.055& 35.55 &                  &   $13.740\pm 0.041$&  $12.687\pm 0.022$&  $12.231\pm 0.020$&  $ 11.914\pm 0.027$\\
   29/10/2004&   308.218& 37.72 & $14.195\pm 0.036$&   $13.846\pm 0.027$&  $12.746\pm 0.017$&  $12.247\pm 0.016$&  $ 11.923\pm 0.021$\\
   30/10/2004&   309.125& 38.62 & $14.268\pm 0.050$&   $13.867\pm 0.013$&  $12.755\pm 0.023$&  $12.248\pm 0.022$&  $ 11.929\pm 0.020$\\
   01/11/2004&   311.132& 40.63 & $14.385\pm 0.033$&   $13.930\pm 0.018$&  $12.767\pm 0.019$&  $12.264\pm 0.008$&  $ 11.928\pm 0.016$\\
   05/11/2004&   315.053& 44.55 & $14.623\pm 0.035$&   $14.046\pm 0.019$&  $12.802\pm 0.016$&  $12.283\pm 0.023$&  $ 11.953\pm 0.031$\\
   11/11/2004&   321.034& 50.53 & $14.903\pm 0.039$&   $14.197\pm 0.018$&  $12.862\pm 0.029$&  $12.282\pm 0.020$&  $ 11.934\pm 0.039$\\
   12/11/2004&   322.067& 51.57 & $14.991\pm 0.034$&   $14.227\pm 0.013$&  $12.848\pm 0.022$&  $12.294\pm 0.016$&  $ 11.934\pm 0.020$\\
   16/11/2004&   326.046& 55.55 & $15.180\pm 0.038$&   $14.329\pm 0.016$&  $12.884\pm 0.018$&  $12.310\pm 0.017$&  $ 11.930\pm 0.014$\\
   20/11/2004&   330.070& 59.57 & $15.347\pm 0.041$&   $14.408\pm 0.024$&  $12.910\pm 0.023$&  $12.301\pm 0.019$&  $ 11.927\pm 0.021$\\
   24/11/2004&   334.072& 63.57 & $15.533\pm 0.041$&   $14.486\pm 0.026$&  $12.942\pm 0.022$&  $12.321\pm 0.022$&  $ 11.946\pm 0.015$\\
   25/11/2004&   335.035& 64.54 & $15.568\pm 0.040$&   $14.499\pm 0.024$&  $12.930\pm 0.028$&  $12.317\pm 0.025$&  $ 11.920\pm 0.019$\\
   04/12/2004&   344.063& 73.56 & $15.861\pm 0.026$&   $14.657\pm 0.016$&  $12.994\pm 0.024$&  $12.352\pm 0.020$&  $ 11.946\pm 0.011$\\
   08/12/2004&   348.041& 77.54 &                  &   $14.726\pm 0.013$&  $13.024\pm 0.014$&  $12.371\pm 0.013$&  $ 11.944\pm 0.019$\\
   12/12/2004&   352.034& 81.53 & $16.160\pm 0.034$&   $14.788\pm 0.020$&  $13.031\pm 0.013$&  $12.372\pm 0.020$&  $ 11.957\pm 0.026$\\
   16/12/2004&   356.096& 85.60 & $16.291\pm 0.034$&   $14.851\pm 0.012$&  $13.075\pm 0.016$&  $12.408\pm 0.018$&  $ 11.978\pm 0.020$\\
   29/12/2004&   369.038& 98.54 &                  &   $15.118\pm 0.012$&  $13.247\pm 0.024$&  $12.524\pm 0.026$&  $ 12.093\pm 0.038$\\
   06/01/2005&   377.044&106.54 &                  &   $15.346\pm 0.011$&  $13.405\pm 0.020$&  $12.651\pm 0.014$&  $ 12.197\pm 0.022$\\
   10/01/2005&   381.044&110.54 & $17.221\pm 0.035$&   $15.522\pm 0.008$&  $13.535\pm 0.013$&  $12.776\pm 0.013$&  $ 12.296\pm 0.011$\\
   13/01/2005&   384.043&113.54 & $17.524\pm 0.040$&   $15.665\pm 0.031$&  $13.660\pm 0.021$&  $12.843\pm 0.022$&  $ 12.374\pm 0.027$\\
   23/01/2005&   394.041&123.54 &                  &   $16.527\pm 0.193$&  $14.488\pm 0.117$&  $13.561\pm 0.114$&  $ 13.005\pm 0.030$\\
   07/03/2005&   436.511&166.01 &                  &   $17.736\pm 0.069$&  $15.822\pm 0.016$&  $14.735\pm 0.021$&  $ 14.154\pm 0.019$\\
   24/03/2005&   454.490&183.99 &                  &   $17.863\pm 0.034$&  $16.002\pm 0.024$&  $14.891\pm 0.022$&  $ 14.349\pm 0.017$\\
   01/04/2005&   462.484&191.98 &                  &                    &  $16.117\pm 0.027$&  $14.990\pm 0.012$&  $ 14.417\pm 0.019$\\
   10/04/2005&   471.473&200.97 &                  &                    &                   &  $15.048\pm 0.027$&                    \\
   13/04/2005&   474.472&203.97 &                  &   $17.961\pm 0.032$&  $16.199\pm 0.013$&  $15.075\pm 0.014$&  $ 14.527\pm 0.013$\\
   15/04/2005&   476.469&205.97 &                  &   $17.997\pm 0.037$&  $16.230\pm 0.027$&  $15.102\pm 0.020$&  $ 14.548\pm 0.020$\\
  \hline														 
  \multicolumn{8}{l}{\rlap{*}\ \* Relative to the epoch of date of explosion(JD = 2453270.5)}				 
  \end{tabular}													 
 \end{minipage}												
  \end{table*}

\setcounter{table}{1}
 \begin{table*}
  \begin{minipage}{126mm}
  \caption{Table 2 contd.}
  \label{tab2}
  \centering
  \begin{tabular}{lccccccc}
 \hline\hline
  Date & J.D. & Phase\rlap{*} & U &  B & V &  R &  I\\
     & 2453000+ & (days) & \\
 \hline
   21/04/2005&   482.403&211.90 &                  &   $18.019\pm 0.035$&  $16.290\pm 0.017$&  $15.152\pm 0.015$&  $ 14.608\pm 0.020$\\
   06/05/2005&   497.392&226.89 &                  &   $18.093\pm 0.009$&  $16.436\pm 0.008$&  $15.314\pm 0.009$&  $ 14.763\pm 0.019$\\
   27/05/2005&   518.448&247.95 &                  &   $18.260\pm 0.037$&  $16.662\pm 0.023$&  $15.551\pm 0.018$&  $ 15.016\pm 0.016$\\
   28/05/2005&   519.385&248.89 &                  &   $18.250\pm 0.018$&  $16.659\pm 0.009$&  $15.538\pm 0.015$&  $ 15.024\pm 0.021$\\
   01/06/2005&   523.393&252.89 &                  &   $18.279\pm 0.019$&  $16.705\pm 0.012$&  $15.575\pm 0.009$&  $ 15.072\pm 0.014$\\
   07/06/2005&   529.402&258.90 &                  &   $18.322\pm 0.017$&  $16.773\pm 0.012$&  $15.652\pm 0.008$&  $ 15.127\pm 0.013$\\
   20/06/2005&   542.419&271.92 &                  &   $18.389\pm 0.019$&  $16.912\pm 0.008$&  $15.792\pm 0.011$&  $ 15.257\pm 0.016$\\
   24/06/2005&   546.410&275.91 & $19.553\pm 0.055$&   $18.414\pm 0.017$&  $16.957\pm 0.017$&  $15.832\pm 0.014$&  $ 15.336\pm 0.011$\\
   25/06/2005&   547.418&276.92 & $19.491\pm 0.055$&   $18.453\pm 0.027$&  $16.975\pm 0.017$&  $15.852\pm 0.016$&  $ 15.363\pm 0.013$\\
   09/07/2005&   561.414&290.91 & $19.630\pm 0.048$&   $18.552\pm 0.020$&  $17.127\pm 0.027$&  $16.018\pm 0.015$&  $ 15.538\pm 0.012$\\
   19/07/2005&   571.406&300.91 &                  &   $18.600\pm 0.017$&  $17.245\pm 0.018$&  $16.140\pm 0.013$&  $ 15.666\pm 0.019$\\
   23/07/2005&   575.267&304.77 & $19.526\pm 0.054$&   $18.615\pm 0.018$&  $17.284\pm 0.010$&  $16.183\pm 0.015$&  $ 15.707\pm 0.020$\\
   01/08/2005&   584.290&313.79 &                  &                    &  $17.371\pm 0.036$&  $16.298\pm 0.008$&  $ 15.856\pm 0.016$\\
   07/08/2005&   590.250&319.75 &                  &   $18.743\pm 0.018$&  $17.446\pm 0.014$&  $16.363\pm 0.014$&  $ 15.921\pm 0.011$\\
   17/08/2005&   600.298&329.80 &                  &   $18.836\pm 0.019$&  $17.571\pm 0.014$&  $16.481\pm 0.015$&  $ 16.059\pm 0.016$\\
   23/08/2005&   606.235&335.73 &                  &   $18.902\pm 0.023$&  $17.638\pm 0.019$&  $16.563\pm 0.012$&  $ 16.125\pm 0.016$\\
   10/09/2005&   624.223&353.72 &                  &   $19.024\pm 0.019$&  $17.831\pm 0.013$&  $16.789\pm 0.011$&  $ 16.371\pm 0.009$\\
   28/09/2005&   642.095&371.59 & $19.982\pm 0.034$&   $19.187\pm 0.018$&  $18.044\pm 0.017$&  $17.024\pm 0.013$&  $ 16.647\pm 0.020$\\
   30/09/2005&   644.133&373.63 & $19.905\pm 0.032$&   $19.182\pm 0.015$&  $18.067\pm 0.009$&  $17.065\pm 0.007$&  $ 16.667\pm 0.010$\\
   17/10/2005&   661.207&390.71 &                  &   $19.278\pm 0.078$&  $18.264\pm 0.018$&  $17.254\pm 0.018$&  $ 16.874\pm 0.027$\\
   27/10/2005&   671.175&400.67 &                  &   $19.422\pm 0.015$&  $18.376\pm 0.010$&  $17.403\pm 0.012$&  $ 17.036\pm 0.014$\\
   23/11/2005&   698.109&427.61 &                  &   $19.680\pm 0.017$&  $18.730\pm 0.016$&  $17.798\pm 0.006$&  $ 17.426\pm 0.017$\\
   26/12/2005&   731.103&460.60 &                  &                    &  $19.202\pm 0.016$&  $18.318\pm 0.013$&  $ 17.919\pm 0.025$\\
   06/03/2006&   801.493&531.29 &                  &                    &                   &  $19.477\pm 0.027$&\\  
   16/03/2006&   811.426&540.93 &                  &                    &  $20.348\pm 0.041$&  $19.608\pm 0.0265$&\\ 

\hline														 
\multicolumn{8}{l}{\rlap{*}\ \* Relative to the epoch of date of explosion(JD = 2453270.5)}				 
\end{tabular}													 
\end{minipage}												
\end{table*}

\subsection{Spectroscopy}
The spectroscopic observations of SN 2004et started on 2004 Oct 16 and continued
until 2005 Dec 30, corresponding to $\sim$ 25 days to $\sim$ 465 days after 
explosion. The journal of observation is given in Table \ref{tab3}. All spectra 
were obtained in the wavelength range 3500-7000 \AA \ and 5200-9200 \AA \ at a 
spectral resolution of $\sim$ 7\AA. A few spectra, in the wavelength range
4000--8500 \AA\ were also obtained with the
1m telescope at the Vainu Bappu Observatory, Kavalur, India,
using the UAG spectrograph. All data were reduced using the standard 
routines within IRAF. The data were bias corrected, flat-fielded and the one 
dimensional spectra were extracted using the optimal extraction method. The 
wavelength calibration was done using FeAr and FeNe lamp spectra. The 
instrumental response curves were obtained using spectrophotometric standards
observed on the same night and the supernova spectra were brought to a relative
flux scale. On a few nights when the spectrophotometric standards were not
observed, the response curves obtained on other nights were used for the flux
calibration. The flux calibrated spectra in the two different regions were 
combined to a weighted mean to give the final spectrum on a relative flux 
scale. The spectra were brought to an absolute flux scale using zero points
obtained by comparing with the photometric magnitudes. The telluric lines are 
not removed from the spectra. 

\begin{table*}
\begin{minipage}{106mm}
\caption{Journal of spectroscopic observations of SN 2004et.}
\label{tab3}
\centering
\begin{tabular}{lrrc}
\hline\hline
Date & J.D. & \multicolumn{1}{c}{Phase\rlap{*}} & Range \\
     & 2453000+ & \multicolumn{1}{c}{(days)} & \AA\ \\

\hline\hline
16/10/04 & 295.1 & 24.60  & 3500-7000; 5200-9200\\
22/10/04 & 301.1 & 30.60  & 3500-7000; 5200-9200\\
27/10/04 & 306.0 & 35.50  & 5200-9200\\
30/10/04 & 309.1 & 38.60  & 3500-7000; 5200-9200\\
01/11/04 & 311.2 & 40.70  & 3500-7000; 5200-9200\\
11/11/04 & 321.0 & 50.50  & 3500-7000; 5200-9200\\
16/11/04 & 326.1 & 55.60  & 3500-7000; 5200-9200\\
18/11/04 & 328.1 & 57.50  & 4000-8500$^{\dagger}$\\
19/11/04 & 329.2 & 58.60  & 4000-8500$^{\dagger}$\\
24/11/04 & 334.0 & 63.50  & 3500-7000; 5200-9200\\
04/12/04 & 344.1 & 73.60  & 3500-7000; 5200-9200\\  
14/12/04 & 354.0 & 83.50  & 3500-7000; 5200-9200\\
17/12/04 & 357.0 & 86.50  & 3500-7000; 5200-9200\\
29/12/04 & 369.1 & 98.60  & 3500-7000; 5200-9200\\
12/01/05 & 383.1 &112.60  & 3500-7000; 5200-9200\\
03/03/05 & 433.5 &163.00  & 3500-7000; 5200-9200\\
13/03/05 & 443.5 &173.00  & 3500-7000; 5200-9200\\
25/03/05 & 455.4 &184.90  & 3500-7000; 5200-9200\\
10/04/05 & 471.4 &200.90  & 3500-7000; 5200-9200\\
21/04/05 & 482.4 &211.90  & 3500-7000; 5200-9200\\
06/05/05 & 497.4 &226.90  & 3500-7000; 5200-9200\\
28/05/05 & 516.3 &245.80  & 3500-7000; 5200-9200\\
01/06/05 & 523.4 &252.90  & 3500-7000; 5200-9200\\
07/06/05 & 529.4 &258.90  & 3500-7000; 5200-9200\\
25/06/05 & 547.4 &276.90  & 3500-7000; 5200-9200\\
19/07/05 & 571.4 &300.90  & 3500-7000; 5200-9200\\
01/08/05 & 584.3 &313.80  & 3500-7000; 5200-9200\\
17/10/05 & 661.2 &390.70  & 3500-7000; 5200-9200\\ 
27/10/05 & 671.1 &400.60  & 3500-7000; 5200-9200\\
23/11/05 & 598.1 &427.60  & 3500-7000; 5200-9200\\
30/12/05 & 735.1 &464.60  & 3500-7000; 5200-9200\\
\hline
\multicolumn{4}{l}{\rlap{*}\ \ Relative to the epoch explosion (JD = 2453270.5).}\\
\multicolumn{2}{l}{{$\dagger$} \ \ Observed from VBO, Kavalur.}\\
\end{tabular}
\end{minipage}
\end{table*}

\subsection{Light Curve}
\begin{figure}
\resizebox{\hsize}{!}{\includegraphics{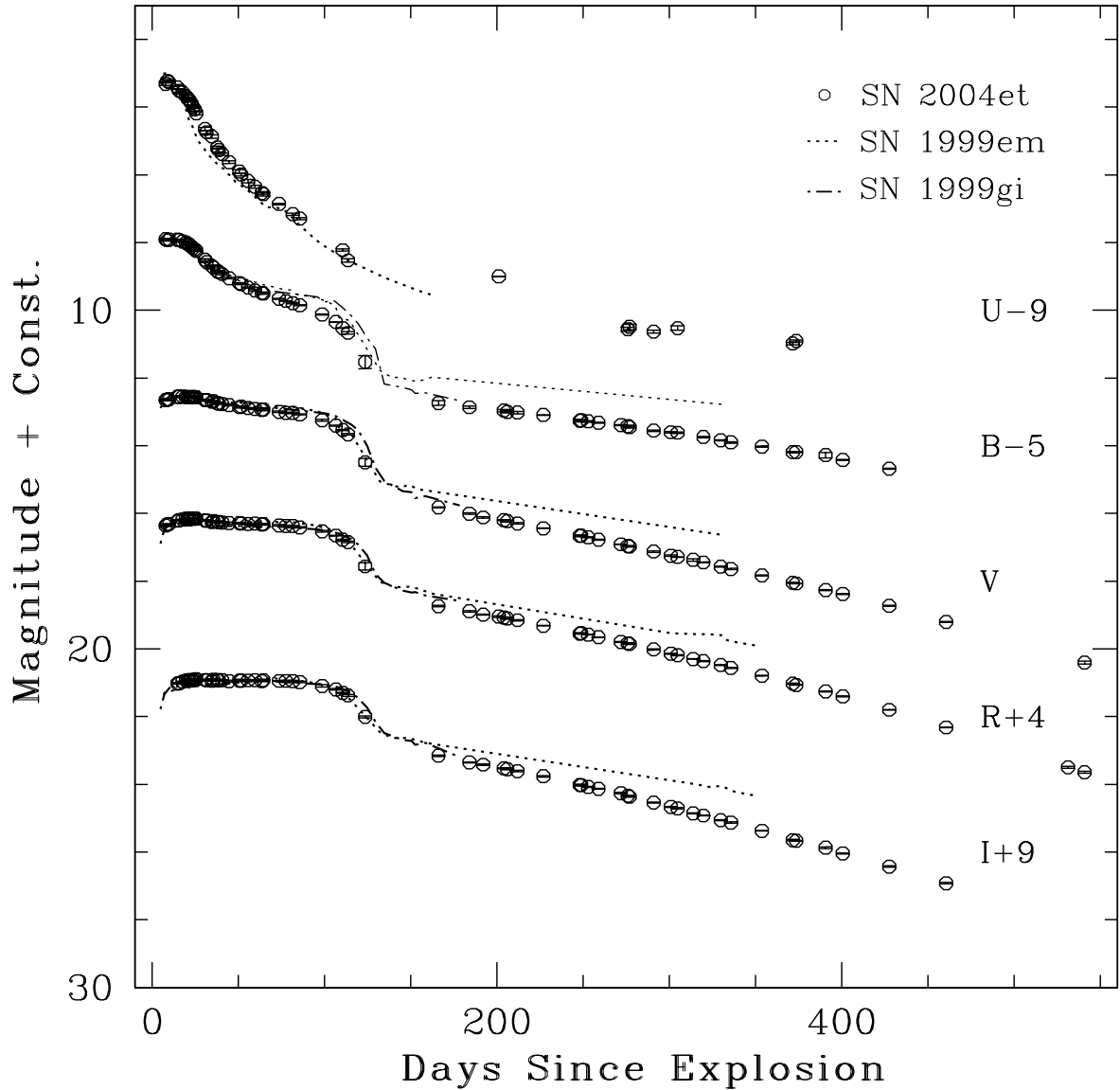}}
\caption[]{$UBVRI$ light curve of SN 2004et plotted with other type IIP SNe. Light curves of SN 2004et have been shifted by the reported amounts, other light curves have been shifted by arbitrary amounts to match those of SN 2004et.}
\label{fig2}
\end{figure}

The light curves of SN 2004et in the $UBVRI$ bands are plotted in Fig. 
\ref{fig2}. Based on the non-detection of the supernova, with a limiting 
magnitude of $19.4\pm1.2$,  on September 22.017 and the subsequent detection at 
a magnitude of $15.17\pm0.16$ on September 22.983, \cite{li1}
constrained the date of explosion to be 2004 September 22.0 (JD 2453270.5).
Our photometric observations of SN 2004et thus span the range of 8 days 
(JD 2453278.2) to 541 days (JD2453811.4) since explosion.

A family of cubic spline fits were made to the observed data points around maximum to estimate the date of maximum and maximum magnitude. The dates of maximum and maximum magnitudes were determined as an average of the values of these fits with  the uncertainties given by the respective standard deviation of the estimates.  The values thus obtained are 
$m_{U}$(max)= 12.17$\pm0.05$ on JD 2453279.93$\pm1.50$, 
$m_{B}$(max)= 12.89$\pm0.02$ on JD 2453280.90$\pm2.13$, 
$m_{V}$(max)= 12.55$\pm0.01$ on JD 2453286.58$\pm0.50$, 
$m_{R}$(max)= 12.15$\pm0.02$ on JD 2453291.47$\pm1.81$ and
$m_{I}$(max)= 11.91$\pm0.03$ on JD 2453294.57$\pm1.37$.
The $B$ maximum occurred $\sim$10 days after the explosion.

The light curves in the $UBVRI$ bands show an initial rise, followed by a 
plateau in the $VRI$ bands, which extends upto $\sim$ 110 days from the date 
of explosion. The plateau in the $VRI$ light curves and a decline rate of 
${\beta}^{B}_{100} = 2.2$ mag in the $B$ light curve over the first 100 days
since maximum light establishes that SN 2004et is a type IIP event, since the
decline rate for SNe IIP is ${\beta}^{B}_{100} < 3.5 $ \citep{patat}.
After the initial rise to maximum, the $U$ band light curve declines rapidly 
till $\sim$ 100 days, while the decline in the $B$ band is less steep compared
to that of the $U$ band. The $V$ band light curve shows a slowly declining 
trend in the plateau phase while the light curve in the $R$ and $I$ bands 
show a nearly constant brightness during the plateau phase. 

Following the plateau phase, the $V$ light curve displays a steep decline of 
$\sim 2$ mag in $\sim 40$ days (day $\sim 110$ to $\sim 150$; JD2453380 to JD2453420). Subsequently,
the decline is linear in all the bands, with decline rates of 
$\gamma_{B} \sim $ 0.64, $\gamma_{V} \sim $ 1.04, $\gamma_{R} \sim $ 1.01 and 
$\gamma_{I} \sim $ 1.07, from $\sim$ 180 days to $\sim$ 310 days after 
explosion. The decline rate steepens  beyond 310 days to
$\gamma_{B} \sim $ 0.84,
$\gamma_{V} \sim $ 1.21, $\gamma_{R} \sim $ 1.31 and $\gamma_{I} \sim $ 1.39.
The photometric evolution of  SNe IIP at late phases is powered by 
radioactive decay of $^{56}$Co into $^{56}$Fe, and the expected decay rate is 
$\gamma=0.98$~mag (100 d)$^{-1}$, especially in the $V$ band \citep{patat}.  
Except for the $B$ band, the decay rates obtained during the early nebular 
phase 
(180--310 days) are close to that of $^{56}$Co decay, suggesting that during 
this phase $\gamma$-ray trapping was efficient. However, the steeper decay 
rate beyond day 310 indicates that either the supernova had become more 
transparent to $\gamma$-rays and $\gamma$-ray leakage was significant, or dust 
formation occurred in the supernova ejecta, or it could be a consequence of 
both the phenomena.  

A comparison of the light curves of SN 2004et with those of the well studied 
SNe IIP, SN 1999em \citep{leonard02}, and SN 1999gi \citep{leonardgi} is made. 
The $UBVRI$ light curves of SN 1999em and SN 1999gi, shifted by arbitrary units 
in magnitude to match the respective light curves of SN 2004et, are also
plotted in Fig. \ref{fig2}. The light curves of all the three SNe are
remarkably similar during the early phases. The length of the plateau is also 
very similar, $\sim $ 100 - 120 days for all three events. However, the rate 
of decline during the nebular phase (beyond day 150) is different; SN 2004et 
has a steeper decline, indicating a probable difference in the fraction of 
$\gamma$-rays that escape through the supernovae ejecta. It may be noted 
 that the late time photometry of the supernovae are affected significantly 
by the underlying background of the host, which may flatten the light curve. However, as
SN 2004et  occurred in the outskirts of the spiral arm, with probably no bright 
underlying region, background contamination in the late phase is expected to be low, but that 
may not be the case with SN 1999em and SN 1999gi.   

\begin{figure}
\resizebox{\hsize}{!}{\includegraphics{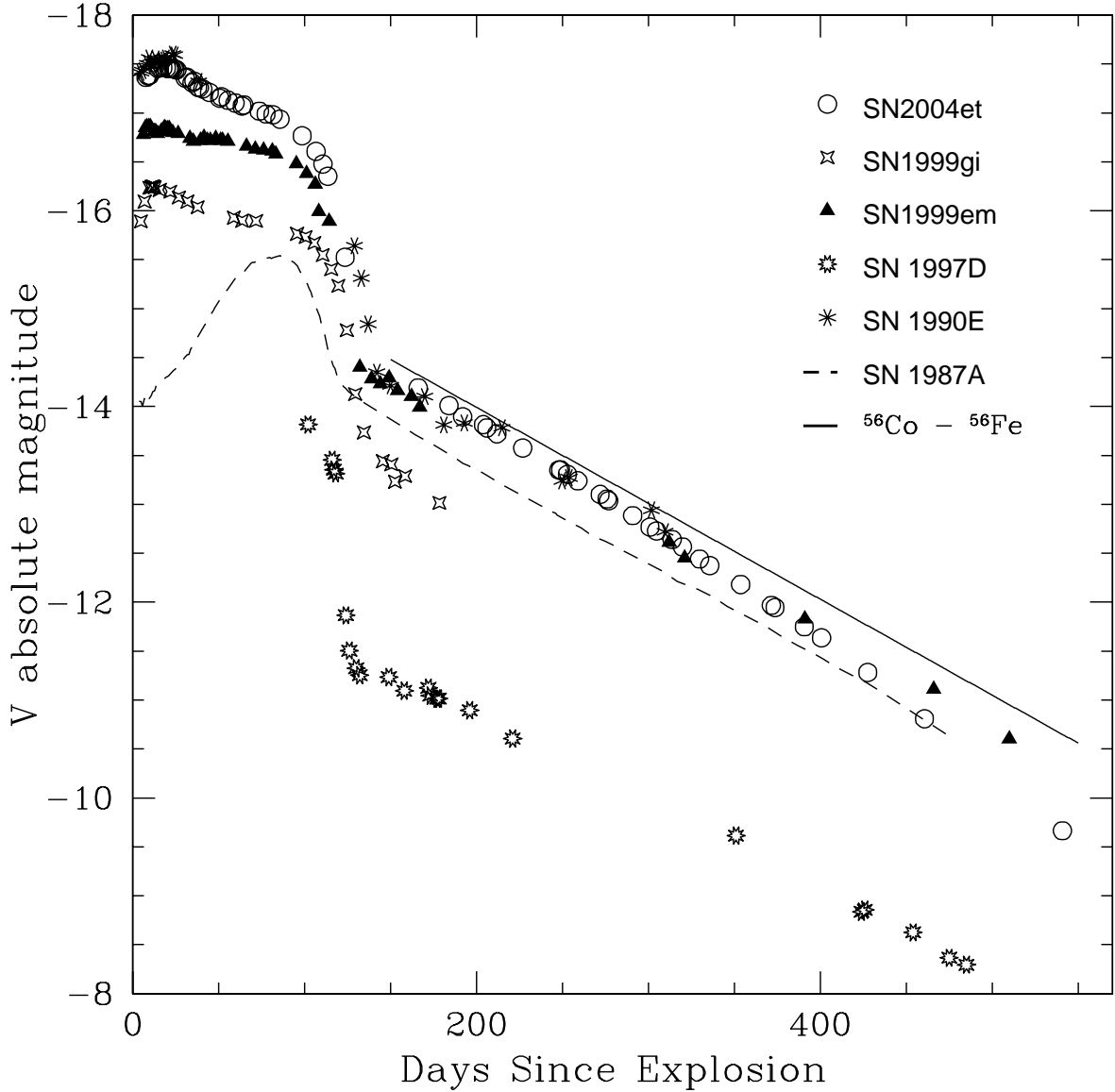}}
\caption[]{Absolute $V$ light curve of  SN 2004et alongwith those of other SNe IIP. The slope of $^{56}$Co to  $^{56}$Fe radioactive decay is also shown. }
\label{fig3}
\end{figure}

In Fig. \ref{fig3}, we compare the absolute $V$ light curve of SN 2004et with 
those of other SNe IIP, namely, SN 1999em (\citealt{leonard02}; 
\citealt{hamuy01}; \citealt{elmhamdi}), SN 1999gi \citep{leonardgi}, SN 1997D 
\citep{benetti2}, SN 1990E \citep{schmidt90e} and SN 1987A \citep{hamuy87a}. 
The supernovae magnitudes have been corrected for extinction using the
\cite{cardelli} extinction law and the following $E(B-V)$: 0.21 for SN 1999gi, 
0.10 for SN 1999em, 0.00 for SN 1997D, 0.50 for SN 1990E and 0.15 for SN 1987A.
A value of $E(B-V)$ = 0.41 is used for SN 2004et (refer Section 4). The 
distance used to calculate the absolute magnitudes and fluxes are 11.1 Mpc for 
SN 1999gi \citep{leonardgi}, 11.7 Mpc for SN 1999em \citep{leonard03}, 13.4 Mpc for 
SN 1997D \citep{benetti2}, 21 Mpc for SN 1990E \citep{schmidt90e1} and 0.0468 Mpc
 for SN 1987A \citep{hamuy87a}. A distance of 5.6 Mpc is used for SN 2004et (refer Section 4).  Distance 
estimates  for SN 1990E and  SN 1999gi are based on the expanding photosphere method (EPM),  
SN 1999em distance estimate is based on Cepheid variables \citep{leonard03}. For
SN 1999em  Cepheid distance is nearly 50\% greater than the values derived using the EPM.
The major source of discrepancy between the distances based on EPM and Cepheid may be attributed to
the underestimate of the theoretically derived dilution factors used in EPM analysis, however, other 
sources of statistical and systematic uncertanity can not be neglected.   

The shape of the light curves of SN 2004et, 
SN 1999em and SN 1999gi are very similar and the length of the plateau is also 
almost the same. The absolute magnitude of SN 2004et is similar to that of
SN 1990E and is brighter than the other SNe IIP  compared here. The average 
absolute $V$ magnitude during the plateau phase, estimated by taking the
unweighted mean of the magnitudes from 20 to 100 days after the explosion is
$-17.14$ for SN 2004et, $-15.97$ for SN 1999gi and $-16.69$ for SN 1999em.

\begin{figure}
\resizebox{\hsize}{!}{\includegraphics{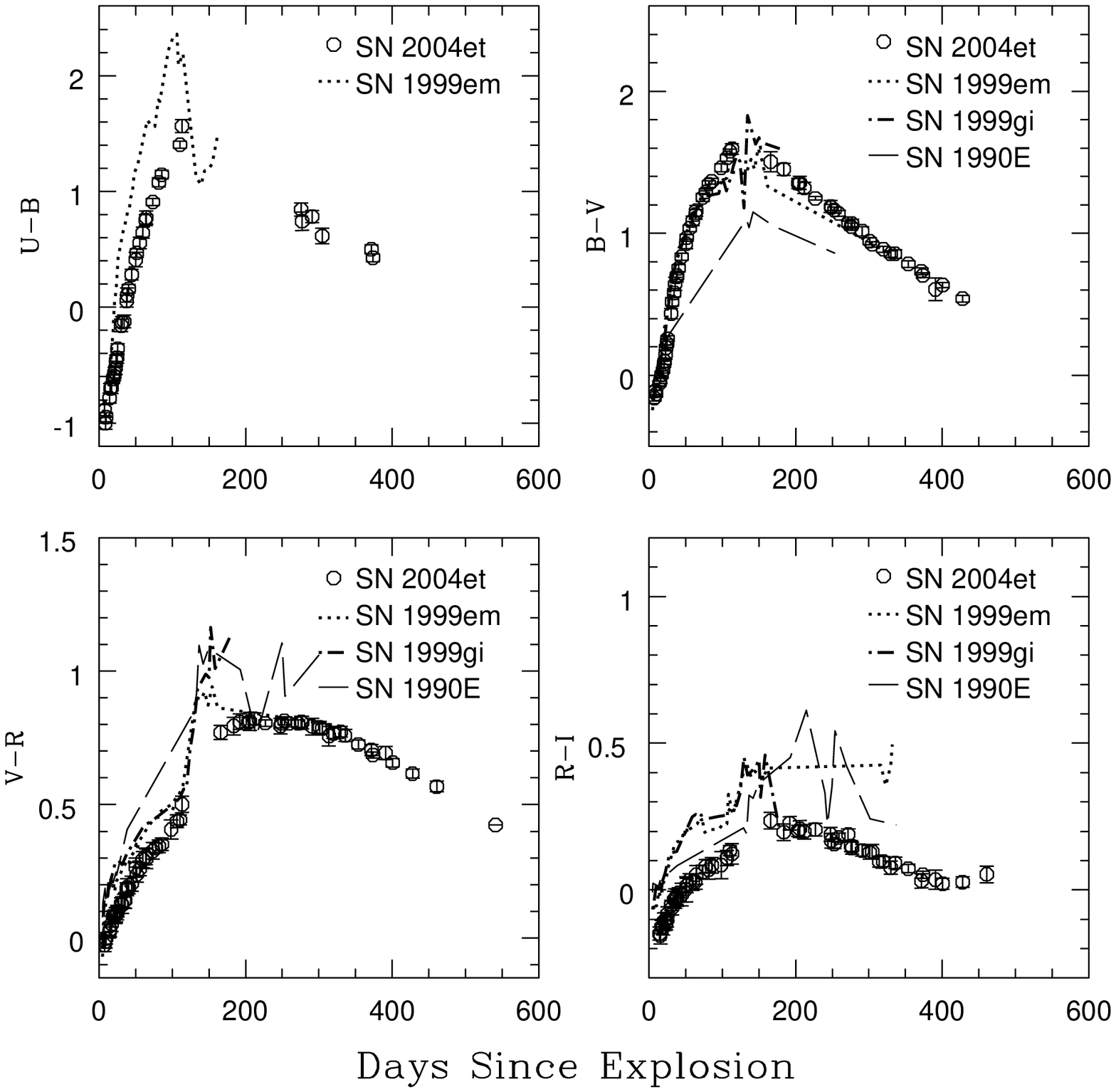}}
\caption[]{Colour curves of  SN 2004et compared with  other SNe IIP}
\label{fig4}
\end{figure}

The reddenning corrected $(U-B)$, $(B-V)$, $(V-R)$ and  $(R-I)$ colour curves of 
SN 2004et are shown in Fig. \ref{fig4}. Also shown in the same figure, for comparison, 
are the respective colour curves of  SN 1999em, SN 1999gi and SN 1990E.
There is a noticable difference in the colour evolution of SN 2004et with those of other 
 SNe IIP in comparison. As noted by  \cite{li1}, during the first month the $(U-B)$ and $(B-V)$ 
colours of SN 2004et evolve more slowly compared to  SN 1999em,  the same trend continues  
till $\sim$ 150 days since explosion.  During this period the overall colours of  
SN 2004et are bluer compared to the other SNe IIP.  Beyond $\sim$ 200 days  after
 explosion, the $(B-V)$ and $(V-R)$ colours of SN 2004et are similar to the other SNe IIP. 
The $(R-I)$ colour is bluer compared to SN 1999em even beyond day 200.

\section{Spectroscopy}
\subsection{Spectroscopic evolution}
\begin{figure}
\resizebox{\hsize}{!}{\includegraphics{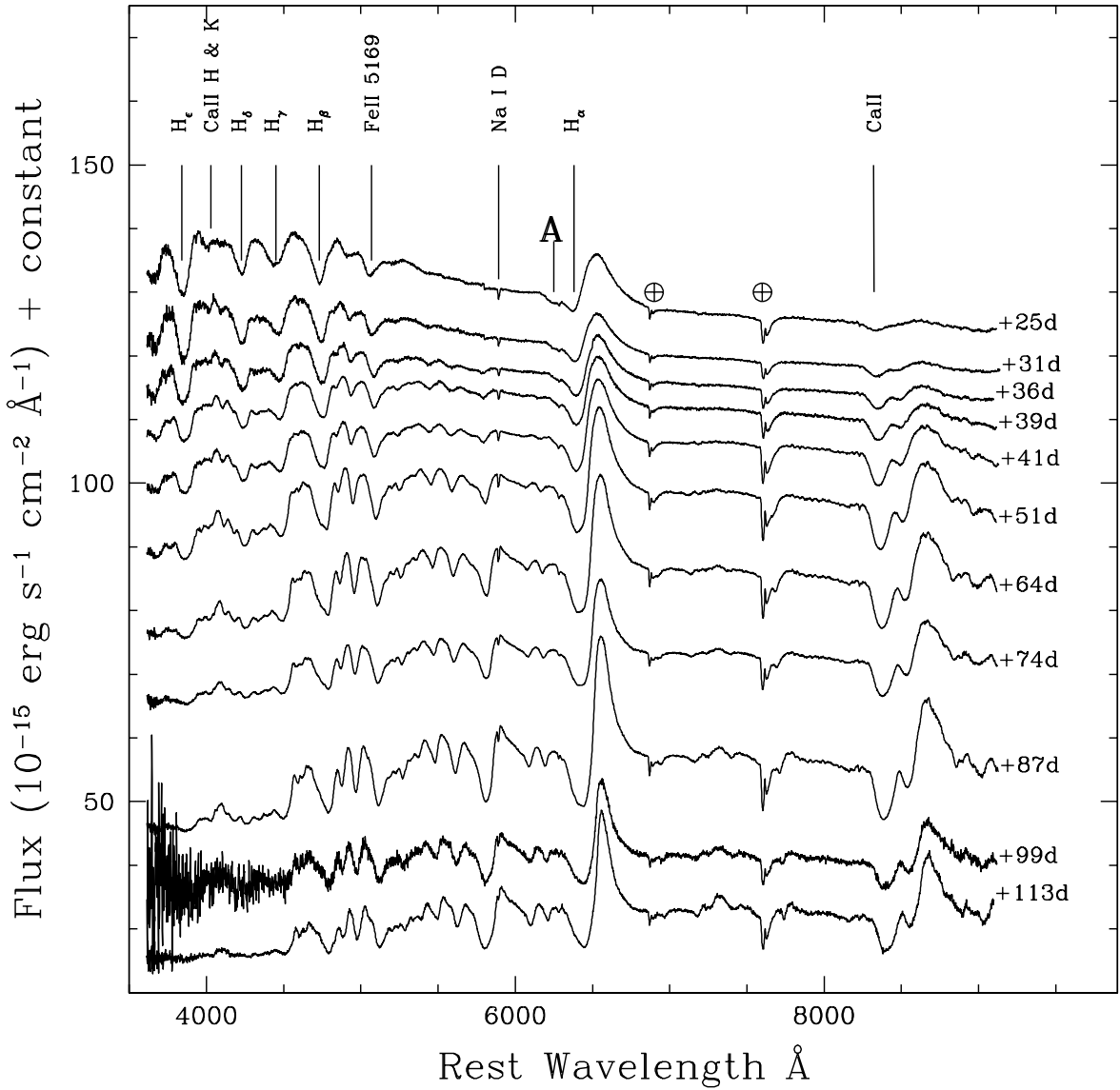}}
\caption[]{Spectral evolution of  SN 2004et during plateau phase.}
\label{fig5}
\end{figure}

\begin{figure}
\resizebox{\hsize}{!}{\includegraphics{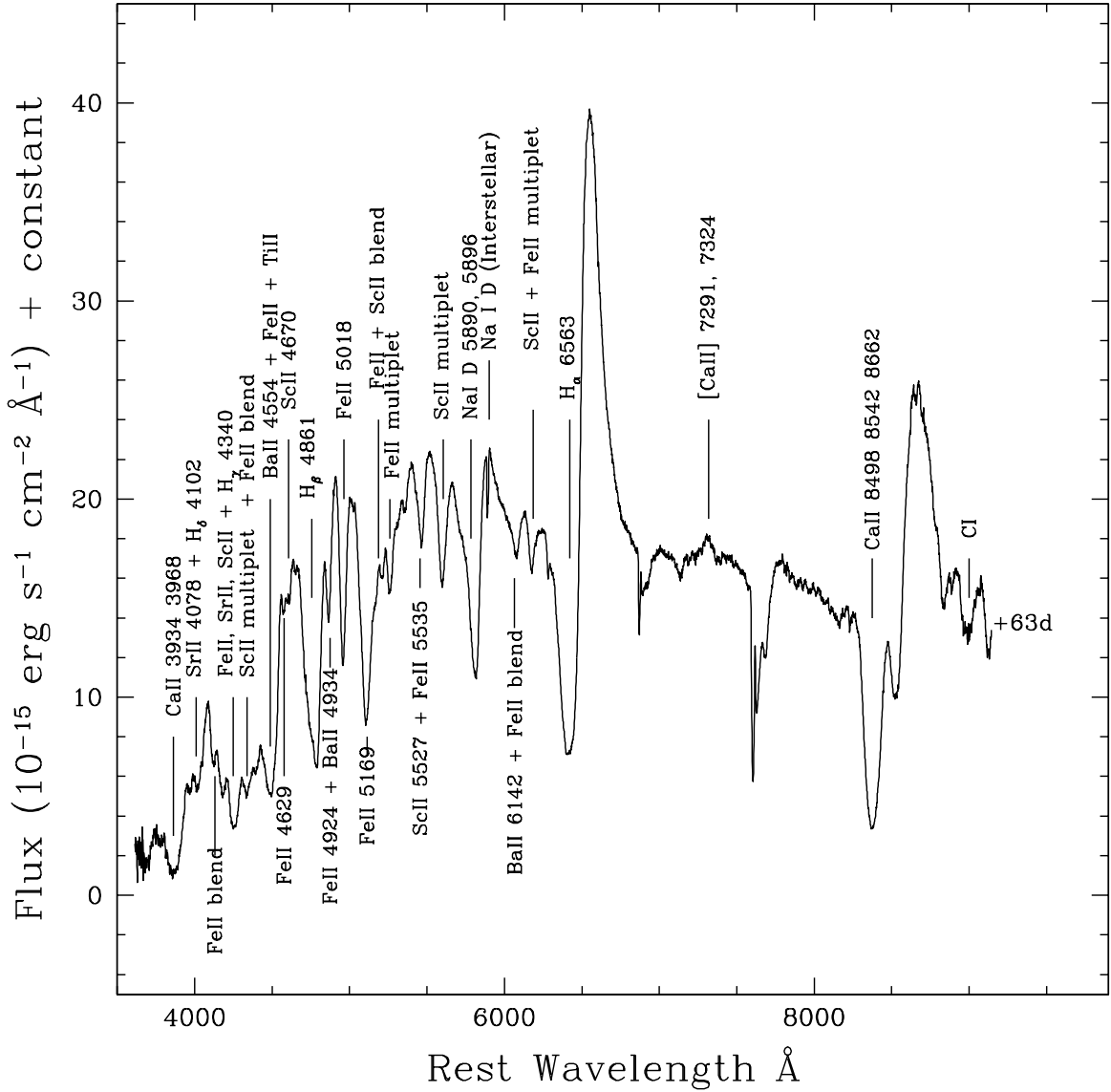}}
\caption[]{Spectral line identification in plateau phase.}
\label{fig6}
\end{figure}

\begin{figure}
\resizebox{\hsize}{!}{\includegraphics{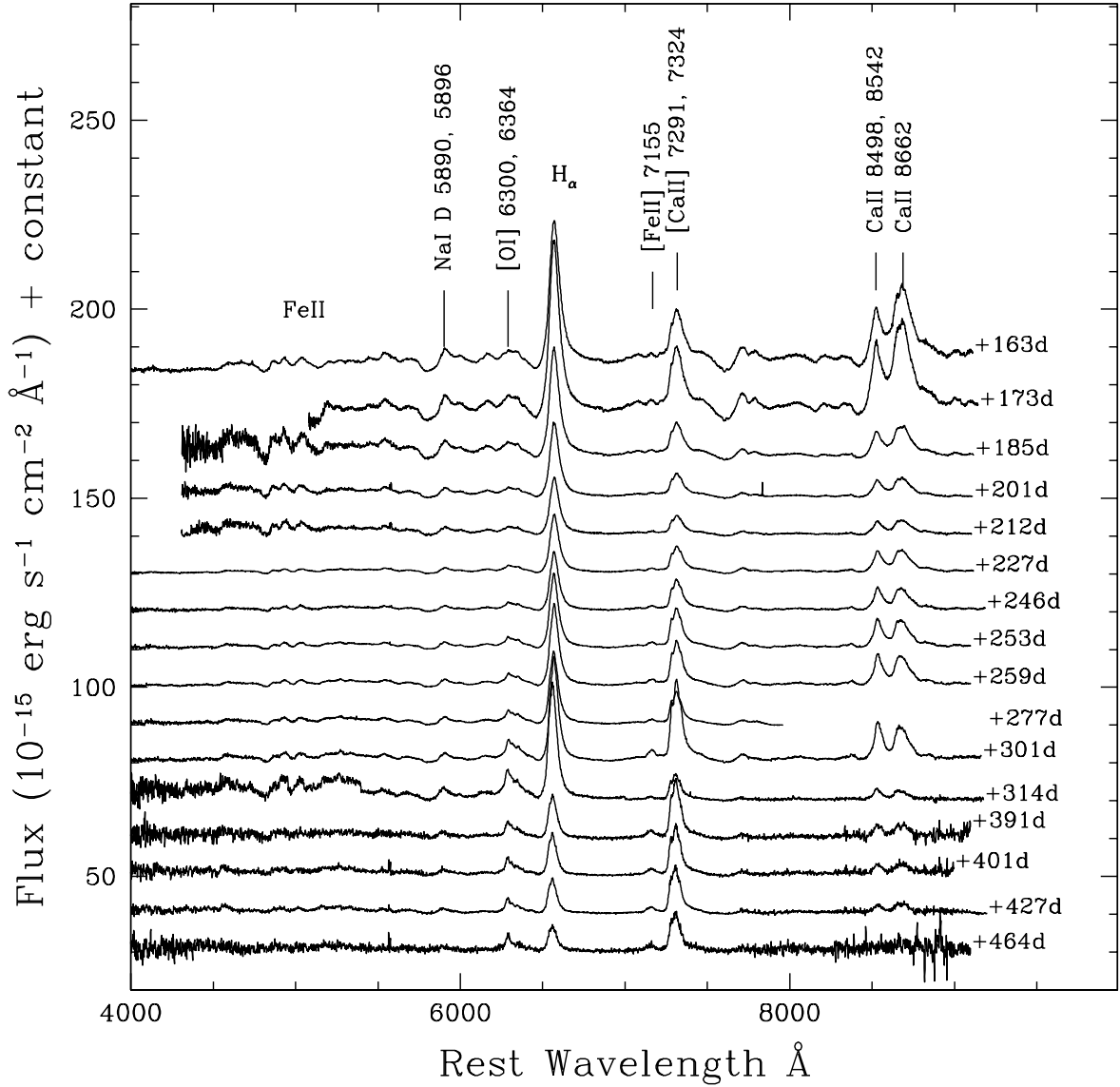}}
\caption[]{Spectral evolution of SN 2004et in the nebular phase.}
\label{fig7}
\end{figure}

The spectroscopic evolution of SN 2004et in the rest frame of the SN is 
displayed in  Fig. \ref{fig5} and Fig. \ref{fig7}. All the spectra have been Doppler corrected for 
the recession velocity of the host galaxy, taken as 45 km sec$^{-1}$ \citep{sandage}.
Our first spectrum corresponds to $\sim$ 25 days after explosion. The spectrum 
has a blue continuum with well developed P-Cygni profiles of hydrogen Balmer lines 
(upto H$\epsilon$).  The H$\alpha$ line is emission dominated with a shallow 
P-Cygni profile, which appears to consist of two components. The second, high velocity 
 absorption component is shown
marked A in Fig. \ref {fig5}.
Apart from the hydrogen Balmer lines, CaII (H \&  K) lines 
3934, 3968~\AA, Fe\,II 5169~\AA\ and the CaII IR triplet 
8498, 8542, 8662~\AA\ are clearly seen in the spectrum. 
Interstellar NaI D absorption features at 5890, 5896~\AA\ are also
present.  By day 39, NaI D absorption lines, Fe\,II lines 
4924, 5018, 5535~\AA, Sc\,II line 5527~\AA\ and other blends of Fe\,II, Sc\,II, Ba\,II 
start appearing. Fig. \ref{fig6} shows the spectrum taken 63 days after the explosion, 
with line identification following  \cite{leonard02}. In the later phase, $\sim 74$ days \
after explosion, except for H$\alpha$ and  H$\beta$, the Balmer lines are obscured by metal lines. 

The two-component P-Cygni profile of H$\alpha$ noted by \citep{li1} in a 
spectrum obtained 20 days after the explosion, is clearly seen in  our 
spectrum of day $\sim 25$. This feature was present until $\sim 113$ days,
until the end of the plateau phase. This feature was also present in the spectrum of 
SN 1999em and has been identified as a high velocity component of H$\alpha$ 
\citep{leonard02}. Similar high velocity component features are noticed in the  
H$\beta$ and  Na I lines also. These high velocity components could be arising due 
to interaction of the supernova ejecta with the pre-supernova circumstellar material.
\cite{barone} interpret the  double component P-Cygni profiles as being  produced 
by a combination of the usual wide P-Cygni profile and a second P-Cygni 
profile with a highly blueshifted absorption, 
corresponding to two line forming regions in the expanding atmosphere of 
the supernova.     

As the supernova ages, the absorption components of H$\alpha$, H$\beta$  
and the CaII IR triplet become narrower and deeper. In the blue region of the 
spectrum numerous  metal lines due to Fe, Sc, Ba and Sr appear and the 
emission line 
strength increases with time till the supernova enters into the nebular phase. The
spectrum obtained $\sim 113$ days after the explosion is still dominated by 
absorption features due to metal lines, indicating that the supernova had not 
yet entered into the nebular phase. However, forbidden lines due to oxygen 
[O\,I] 6300, 6364~\AA,  iron [Fe\,II] 7155~\AA, calcium [Ca\,II] 
7291, 7324~\AA\  start appearing in the spectrum. The spectrum 
obtained $\sim 163$ days after explosion shows a significant evolution 
compared to  day 113. The spectrum is more emission dominated, 
featuring the onset of the nebular phase. In this spectrum, the H$\alpha$ 
profile has narrowed down and the nebular emission lines of [O\,I] 6300, 
6364~\AA\  and [Ca\,II] 7291, 7324~\AA\  are well developed. 
  
\begin{figure}
\resizebox{\hsize}{!}{\includegraphics{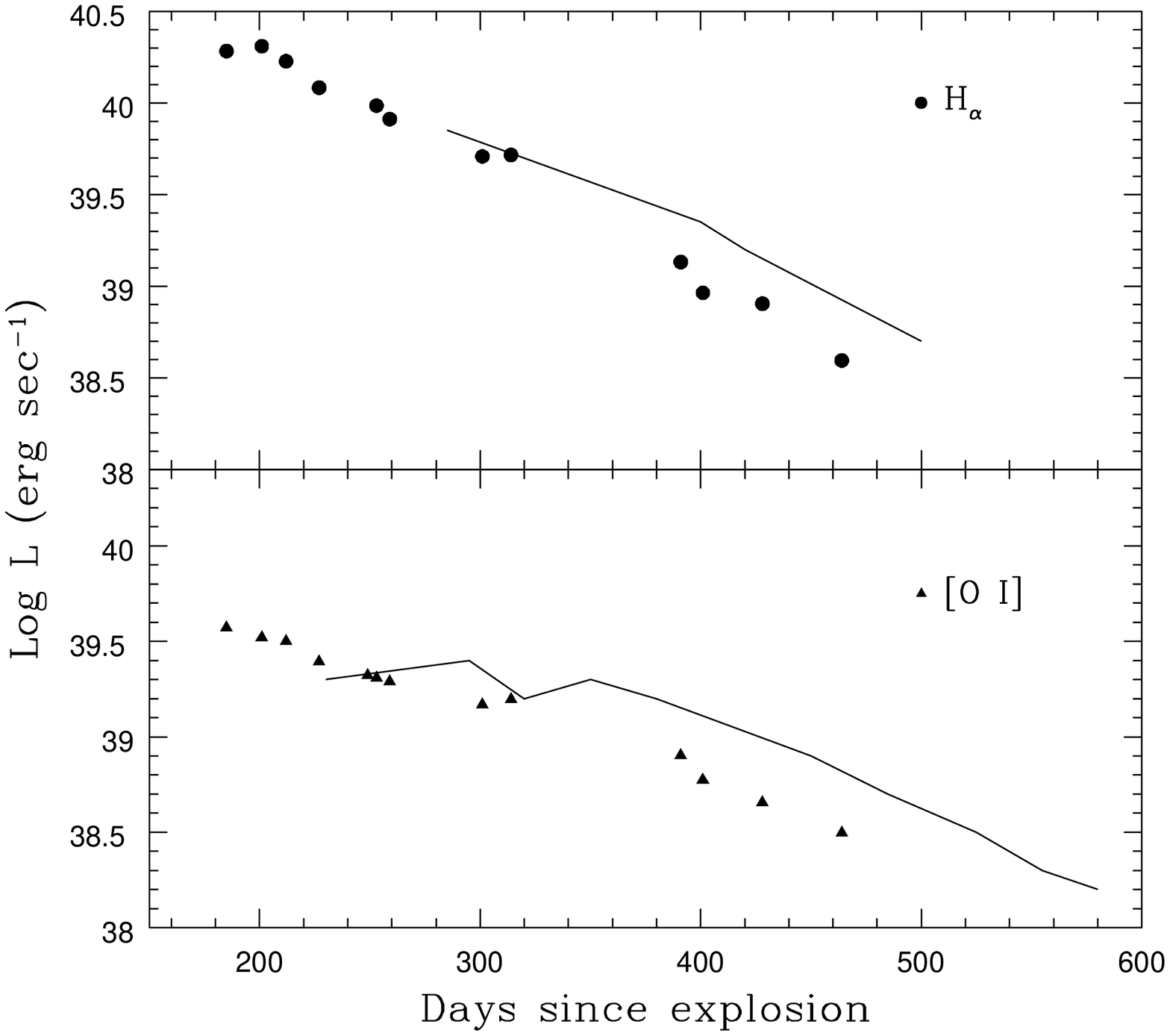}}
\caption[]{The temporal evolution of the luminosity of  nebular lines H$\alpha$ and [OI] for 
SN 2004et, overplotted curve corresponds to SN 1987A.}
\label{fig8}
\end{figure}

In the spectra of day 391 and later, the [Ca\,II] 7291, 
7324~\AA\ line is strong compared to H$\alpha$, a feature noticed in the spectrum 
of SN 1999em around the same epoch \citep{elmhamdi}, but  not detected in 
the SNe IIP, SN 1990E \citep{benetti1}, SN 1997D \citep{benetti2} and
SN 2003gd \citep{hendry}. Fig. \ref{fig8} shows the luminosity evolution of
the nebular lines, H$\alpha$ and [OI] 6300, 6364 \AA\ for SN 2004et. Also
shown in the same figure, for a comparison,  are the corresponding luminosities
for SN 1987A at similar epochs. The nebular line luminosities of SN 2004et
before day 300 are similar to SN 1987A. Beyond day 400, SN 2004et lines have
lower luminosities.

\subsection{Expansion velocity of SN 2004et}
\begin{figure}
\resizebox{\hsize}{!}{\includegraphics{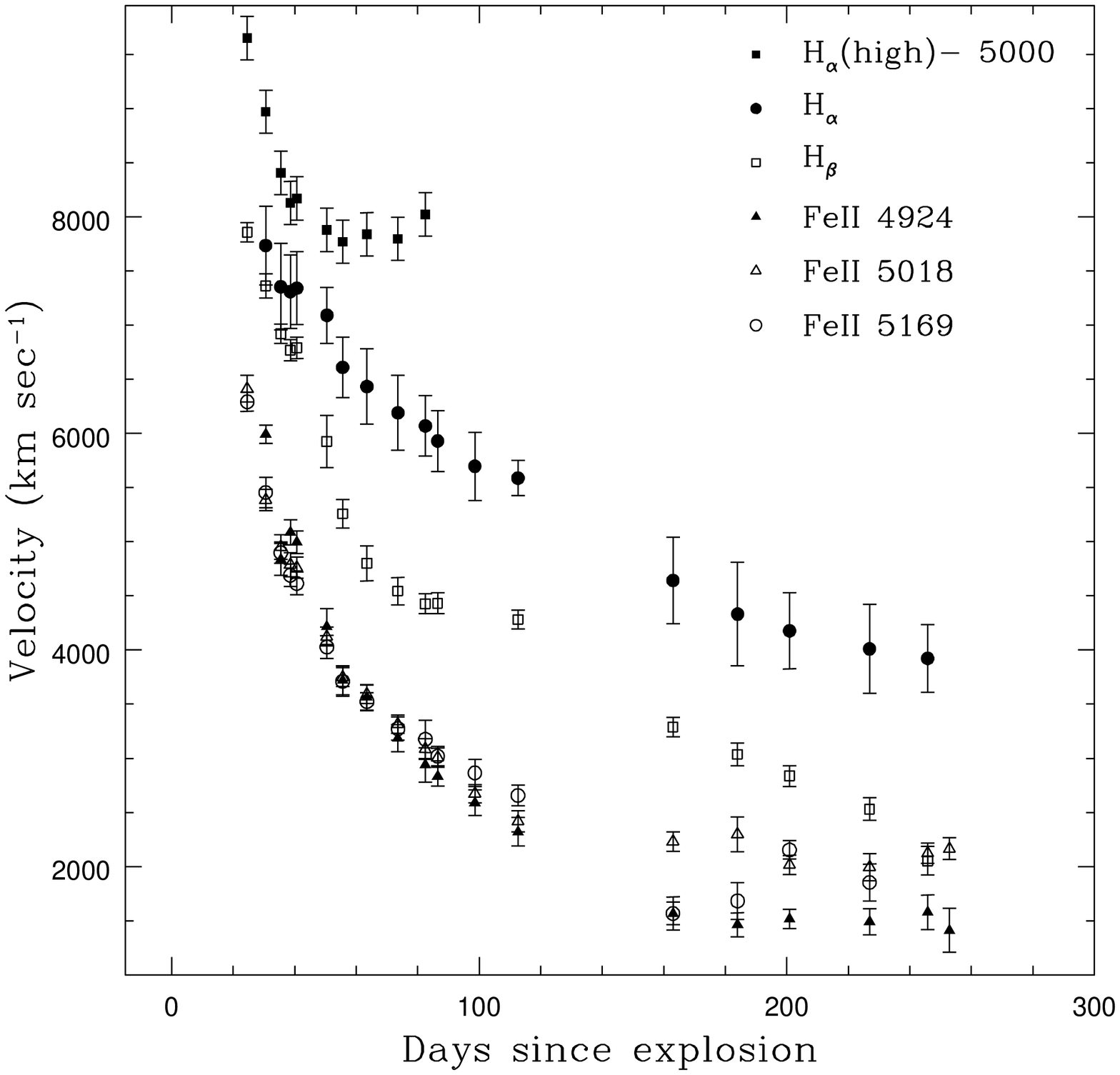}}
\caption[]{Velocity evolution of SN 2004et}
\label{fig9}
\end{figure}

The photospheric velocity of the ejecta can be estimated from the minimum of the absorption 
feature of weak, unblended lines. \cite{hamuy01} have shown that the use of
Fe\,II (multiplet 42) at  4924, 5018, 5169~\AA\ provides a good
estimate of the photospheric velocity during the plateau phase. 
As the supernova ages, it becomes difficult to
define the minimum of Fe\,II 5169~\AA\ line and hence the velocity 
measurement using this line is not a good estimate at these phases. 
The photospheric velocity estimated using these lines is plotted in Fig. 
\ref{fig9}. The velocity estimated using the Fe\,II 4924~\AA\ line 
(ref. Fig. \ref{fig9}) is lower due to the blending of this line with Ba\,II
4934~\AA\ \citep{hendry}. 

Also plotted in the Fig. \ref{fig9} are the velocities determined using the absorption minima 
of H$\alpha$, H$\beta$, and the high velocity component of H$\alpha$. These 
velocity estimates are higher due to higher optical depths. The evolution of 
the high velocity component of H$\alpha$ is similar to that of normal component.
A comparison of the photospheric expansion velocity  determined for SN 2004et,
using the weak unblended iron lines, with those of SN 1999em \citep{hamuy01} 
and SN 1999gi \citep{leonardgi} indicates that SN 2004et has a higher expansion 
velocity at simliar epochs. For further analyses, we use the average 
photospheric velocity estimated using the weak iron lines.

\section{Reddening and distance estimates}	
The interstellar Na lines are clearly seen in the spectra presented here. 
The equivalent width of the Na I D lines measured from the spectra presented
here is 1.70\AA, which corresponds to a total reddening of $E(B-V) = 0.43$ mag, 
following the empirical relation by  \cite{barbon}. This estimate is consistent with
the estimate by \cite{zwitter} based on equivalent widths of the Na I lines in 
a high resolution spectrum. The Galactic reddening in the direction of NGC 6946 
is $E(B-V) = 0.34$ mag \citep{schlegel}, which implies that the supernova has 
suffered minimal extinction in the host galaxy. We use $E(B-V) = 0.41$ \citep{zwitter}
 mag as the total reddening for further analysis. 

The HI Tully Fisher relation \citep{pierce} indicates a distance of 5.5 Mpc
to NGC 6946,
while the CO Tully Fisher relation \citep{schoniger} indicates a distance of
5.4 Mpc. The ``expanding photosphere method'' for the IIP supernova SN 1980K that 
occurred in NGC 6946 \citep{schmidt94} yields a distance of 5.7 Mpc. 
\cite{hamuy02} have shown that the expansion velocities of the ejecta of
SNe IIP are correlated with their bolometric luminosities during the plateau
phase. Based on this correlation they establish a ``standard candle method (SCM)'' 
to estimate the distance to the supernovae. \cite{nugent}  proposed a refinement to the SCM method 
by introducing the use of $(V-I)$ colour during the plateau phase at day 50 to perform extinction 
correction rather than relying on colours at the end of the plateau. Using equation 1 of 
\cite{nugent}, with the fitting parameters for low redshift  SNe IIP and the observed expansion velocity at
day 50  $v_{50} = 4122\pm170$ km sec$^{-1}$ we estimate the distance to 
SN 2004et as $5.7^{+0.3}_{-0.3}$  Mpc.  Based on all the estimates, an average distance
 of 5.6 Mpc is assumed for SN 2004et.   

\section{$UBVRI$ bolometric light curve and $^{56}$Ni mass}
\subsection{$UBVRI$ bolometric light curve}
\begin{figure}
\resizebox{\hsize}{!}{\includegraphics{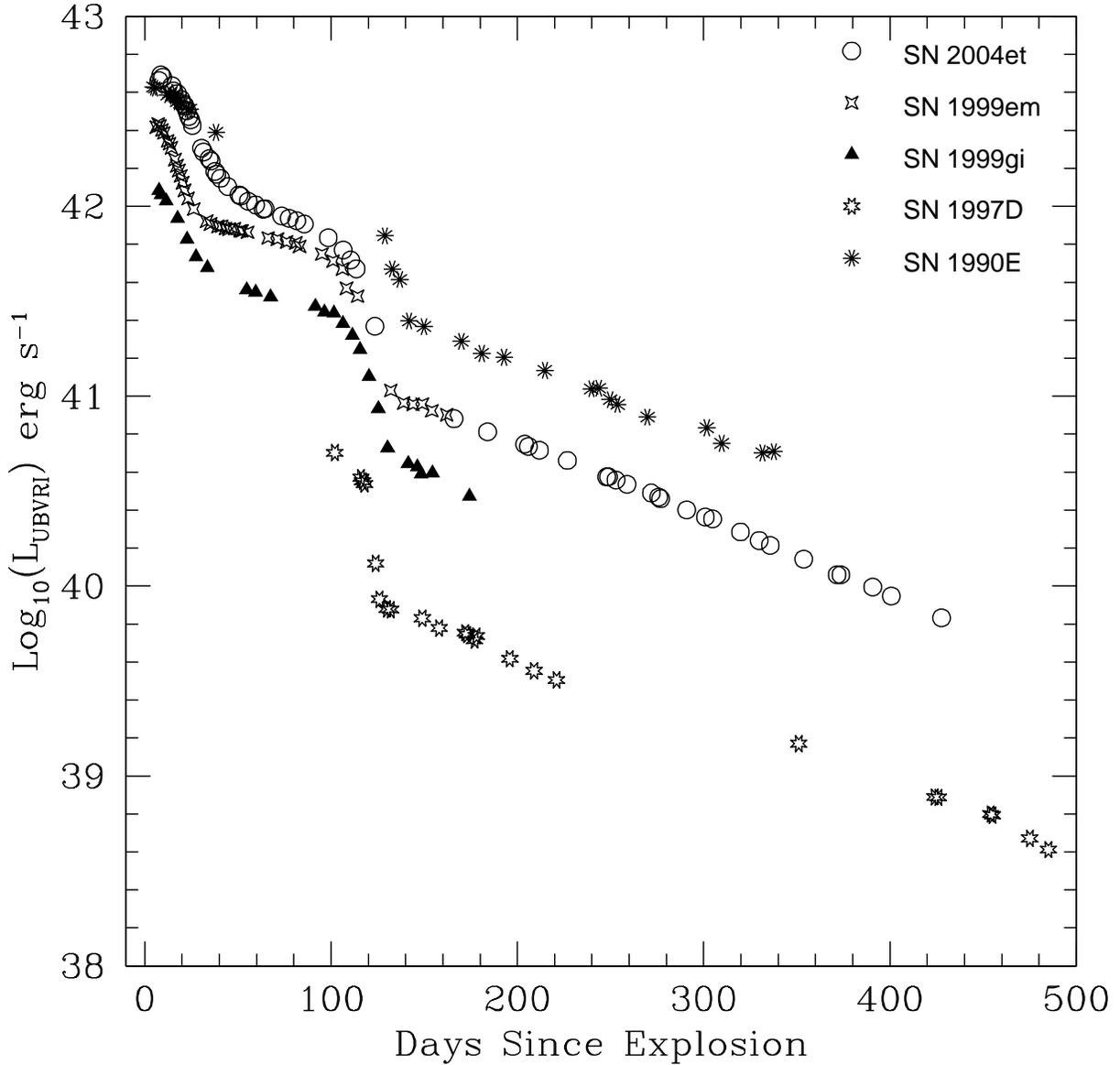}}
\caption[]{Comparison of $UBVRI$ bolometric light curve of  SN 2004et with those of other  type IIP SNe.}
\label{fig10}
\end{figure}

The $UBVRI$ photometry presented in Section 2.3  is used to derive the 
$UBVRI$ bolometric light curve of SN 2004et. The optical magnitudes have been  corrected for  
reddening values mentioned in Section 2.3  using \cite{cardelli} extinction law, the corrected 
magnitudes are then   converted to  fluxes according to \cite{bessel}. 
The $UBVRI$ bolometric fluxes were derived by fitting a spline curve to the $U, B, V, R$ and $I$ 
fluxes and  integrating it over the wavelength range 3200\AA \ to 10600\AA,  determined by the 
response of the filters used.  There are some gaps in the $U$ band light curve specially 
after the plateau; the missing magnitudes were obtained by interpolation. In  
  the later stage, however,  the light curve is linearly extrapolated to get the missing magnitudes
 in the $U$ band. No corrections have been  applied for the missing fluxes in the ultra-voilet 
and the near infra-red region.  Fig. \ref{fig10} shows the $UBVRI$ bolometric light curve of SN 2004et. 
Also plotted in the same figure are the $UBVRI$ bolometric light curves of  
SN 1999em, SN 1999gi, SN 1997D and SN 1990E. The $UBVRI$ bolometric light curve for  SN 1990E is
  taken from the literature \citep{schmidt90e}, while for 
other supernovae the $UBVRI$ bolometric luminosities are derived from the photometry reported in the 
references cited in Section 2.3.  Since no $U$ band magnitudes were available for SN 1999gi and 
SN1997D the contribution of the $U$ flux to the total $UBVRI$ flux was assumed to be similar 
to SN 1999em and SN 2004et.
A comparison of the $UBVRI$ bolometric light curve of SN 2004et with the other SNe IIP
indicates SN 2004et had a higher luminosity. The luminosity during the
initial phases is very similar to SN 1990E but its considerably lower  in the post-plateau decline.

\subsection{$^{56}$Ni mass}
For most SNe IIP, the early (150-300d) bolometric luminosity on the radioactive
tail is equal to the luminosity of the radiaoctive decay of $^{56}$Co.
The mass of $^{56}$Ni synthesized during the supernova explosion can thus be 
estimated from the late time bolometric light curve. In absence of infra-red 
photometry for SN 2004et, we followed \cite{hamuy03},
to estimate the tail bolometric luminosity $L_t$, using the $V$ magnitudes during the
nebular phase and a bolometric correction of 0.26 mag. The tail luminosity
is then used to estimate the nickel mass using equation 2 in \cite{hamuy03}.
Based on the luminosity of SN 2004et during 250 to 315 days, 
estimated using the $V$ magnitudes, we 
estimate the $^{56}$Ni mass to be $0.060\pm0.02\,\, M_\odot$.

\cite{elmhamdia} find a correlation with the $^{56}$Ni mass and the rate of 
decline in the $V$ band light curve from the plateau to the tail. The maximum 
gradient during the transition from plateau to nebular phase, defined by a 
steepness parameter $S=dM_V/dt$, is found to 
anticorrelate with $^{56}$Ni mass, in the sense that the steeper the decline 
at the inflection, the lower is the $^{56}$Ni mass. 
Using a sample of 10 SNe IIP, \cite{elmhamdia} derive the following
relation
\begin{center}
\begin{equation}
log \ \Big(\frac {M_{Ni}}{M_\odot}\Big) = -6.2295 \ S - 0.8147 
\end{equation}
\end{center}
 
An accurate determination of the steepness parameter $S$ requires a well 
sampled $V$ band light curve during the end of plateau to the beginning of the 
radioactive tail. Unfortunately, we do not have a very well sampled light
curve during this phase. However, with the available points, we estimate the
steepness parameter $S$ as $0.062\pm0.02$. Using this, the mass of 
$^{56}$Ni is found to be $0.062\pm0.02 {M_\odot}$, which agrees 
very well with the $^{56}$Ni mass derived from $V$ magnitude on the radioactive 
tail.

The nickel mass may also be estimated comparing the bolometric light curve of 
SN 2004et with that of SN 1987A, assuming that the $\gamma$-ray deposition for 
SN 2004et is the same as that for SN 1987A. The tail bolometric luminosity of
SN 2004et ($\sim 250-300$) is found to be $\sim 1.6$ times fainter than that 
of SN 1987A. This implies a $^{56}$Ni mass of $0.048\pm 0.01 {M_\odot}$ for 
SN 2004et, for a value of $0.075\, M_\odot$ for SN 1987A \citep{turatto}. This will be the 
lower limit of  $^{56}$Ni mass, as the bolometric curve for SN 2004et does not include 
contribution from near-infra-red region.

The mean nickel mass derived using the $V$ band magnitudes in the nebular phase  
and steepness parameter $S$ is $0.06\pm 0.02\, M_\odot$.

\cite{chugai} and \cite{elmhamdia} find that the H$\alpha$ luminosity at the 
nebular phase can be used to estimate the nickel mass. The H$\alpha$ luminosity
is found to be proportional to $^{56}$Ni mass during 200-400 days after the
explosion. Comparing the H$\alpha$ luminosity of SN 2004et around day 250
(ref. Fig. \ref{fig8} top panel) with that of SN 1987A during the same 
phase, it
is found that the H$\alpha$ luminosities in both SNe are similar. Assuming the
mass, energy and mixing conditions do not vary strongly, this indicates the
nickel mass in SN 2004et to be similar to that estimated for SN 1987A.
This is consistent with the photometric estimates. 

\section{Dust formation in the ejecta}

The formation of dust in the supernova 
ejecta  increases  the rate of decline of the optical, especially the $V$ band, 
light curve, as the optical light is reprocessed by the dust and an excess 
emission in $IR$ is observed. Further, dust formation in the inner envelope of 
the supernova ejecta shifts the peak of the optical and infrared emission lines 
towards blue, due to a preferential extinction of the redshifted edge of the  
emission lines by the dust \citep{lucy}. The signature of dust formation was
seen in SN 1987A \citep{danziger} and SN 1999em \citep{elmhamdi} as an 
observable blueshift in the emission peaks of the [OI] doublet components,
beyond $\sim 400$ days. It should be noted that both SNe showed an early
blueshift ($\sim 200-300$ days) in the [OI] component, probably as a result of
superposition of the blend of Fe II lines (multiplet 74).

\begin{figure}
\resizebox{\hsize}{!}{\includegraphics{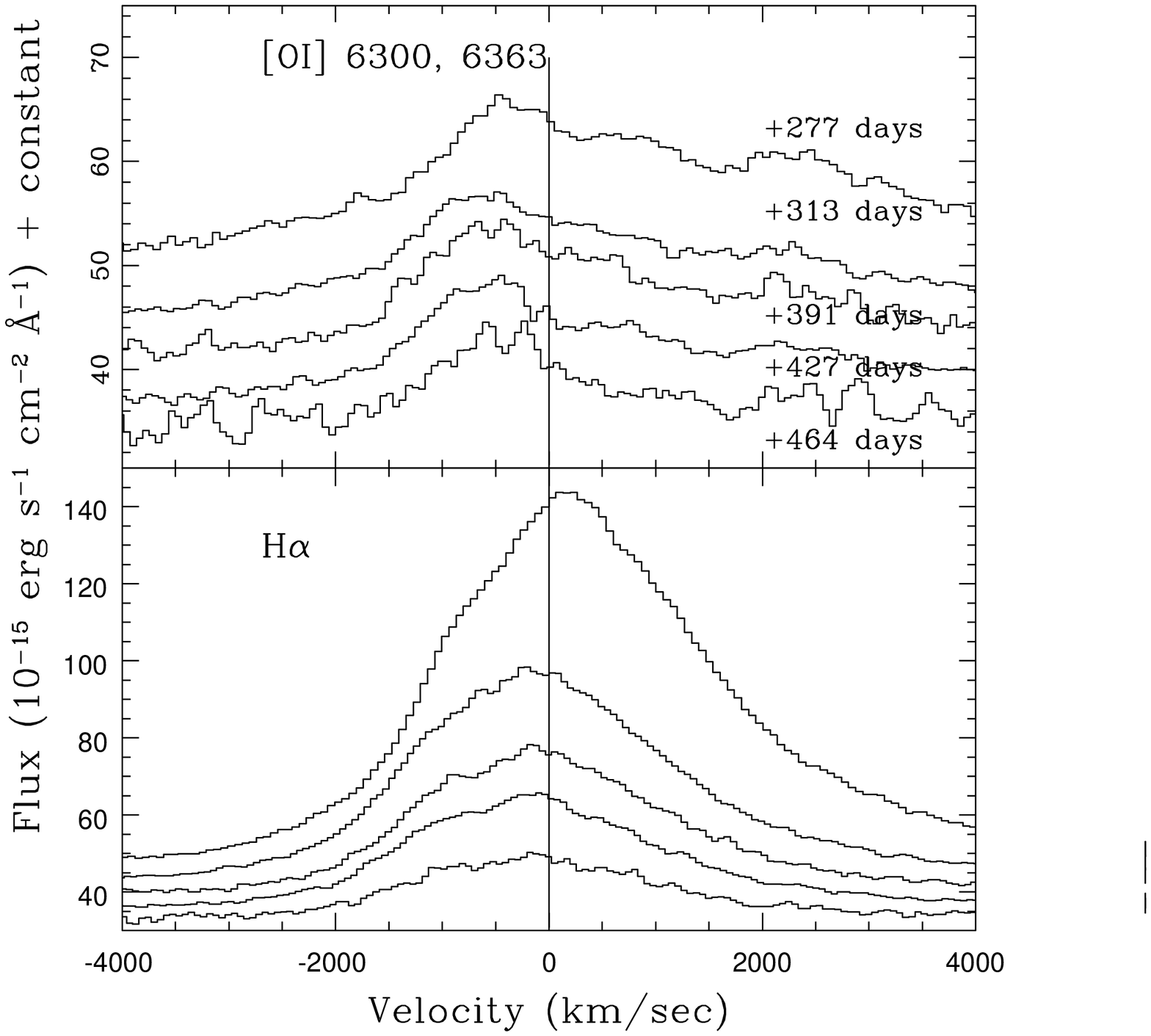}}
\caption[]{Temporal evolution of line profile of  oxygen doublet [OI] 6300, 6364
\AA\ (top panel) and H$\alpha$ (bottom panel). The vertical line
corresponds to the zero velocity of the [OI] 6300 \AA\ and H$\alpha$.
 The epochs are as marked for the [OI] profile.}
\label{fig11}
\end{figure}

Fig. \ref{fig11} shows the temporal evolution of the H$\alpha$ and [O\,I] 
6300, 6364 \AA\ line profiles during 277--465 days. A blueshift in the
emission peak is clearly seen beyond day 300  in both lines. Further,
the H$\alpha$ line shows a clear flattening of the emission peak. 
The blueshift in the emission peak and the flattening were seen in both 
SN 1987A 
and SN 1999em during the dust formation epoch. The absence of
any `blue bump' in the [OI] line due to Fe II 6250 \AA\ feature (ref. Fig.
\ref{fig11}, top panel) indicates that the observed blue shift in the emission 
peak is not due to a blending of Fe II lines, and more likely due to dust 
formation. 

It is evident from Fig. \ref{fig3} that the decline rate of the light curve of SN 2004et 
in the $V$ band starts steepening $\sim 320$ days after the explosion. A
similar effect was seen in both SN 1987A and SN 1999em during dust formation.
It should, however, be noted here that dust formation occurred at much later
phases, beyond day 400 in both these SNe.

The evolution of the emission line profiles together with the steeping of the 
light curve beyond day $\sim 320$ suggest an early dust formation in the case of SN 2004et.

\section{Progenitor star properties}

The supernova outburst properties depend on three basic parameters: the 
mass $M$ of the envelope that is ejected, the radius $R$ of the star prior 
to the outburst and the energy $E$ of explosion. 
An analysis of supernova light curve combined with the spectroscopic evolution 
allows the derivation of these important parameters. 
With the help of hydrodynamic models of SNe IIP, \cite{litvinova} derived 
expressions for these parameters in terms of the observable parameters, 
namely, the length of the plateau $\Delta t$, the absolute magnitude $M_V$ at 
the midpoint of the plateau and the velocity of photosphere $U_{ph}$ at 
mid plateau.  

The light curve of SN 2004et indicates the length of the plateau to
be $\Delta t$ = 120$\pm10$ days. The absolute magnitude at mid plateau is
$M_V = -17.14$~mag. The weak iron lines indicate a mid plateau velocity 
$U_{ph}= 3560\pm 100$ km sec$^{-1}$. Using these values and the relations 
given by Litvinova \& Nadyozhin  (1985) we estimate the explosion energy 
$E_{\rm{exp}}= {1.20}^{+0.4}_{-0.3} \times 10^{51}$ ergs. 

The luminosity of [OI] doublet $\sim 1$~yr after the explosion is powered 
by the $\gamma$-ray deposition and by ultraviolet emission arising from the
deposition of $\gamma$-rays in oxygen-poor material. The [OI] doublet
luminosity is related to the mass of oxygen, the `excited' mass in which the
bulk of the radioactive energy is deposited and the efficieny of transformation
of the energy deposited in oxygen into the [OI] luminosity. The [OI] luminosity
in SN 1987A implies an oxygen mass in the range $1.5-2\,M_\odot$. 
The fact that derived [OI] luminosity for SN 2004et before dust formation is 
comparable to that of SN 1987A at similar epochs implies similar oxygen mass in SN 2004et also.
The nucleosynthesis computations (\citealt{chugai1}, \citealt{woosley}) 
indicate that this oxygen mass corresponds to a main-sequence stellar mass of 
$20\, M_\odot$. An analysis of the radio light curve of SN 2004et by \cite{chevalier}
indicates a mass loss rate for the progenitor that suggests a mass of $20\, M_\odot$, similar to
the estimate based on a comparison of the [OI] luminosity with that of SN 1987A.
However, \cite{li1} estimate the progenitor mass to be $\sim \ 15\, M_\odot$ and also 
the progenitor to be a yellow supergiant.

\section {Summary}

We have presented $UBVRI$ photometric and spectroscopic data for 
SN 2004et from $\sim 8$ until $\sim 541$ days after the explosion.
The shape of the light curve as well as the spectral evolution indicate that
it is Type IIP supernova, which was caught young, soon after the shock breakout.
SN 2004et reached maximum in $B$ band on JD 2453280.90, $\sim 10$ days after 
the explosion. The luminosity at maximum indicates SN 2004et to be at the
brighter end of SNe IIP.
The late phase photometry indicates that the decline rate of light 
curve during $\sim 180-300$~d was similar to the $^{56}$Co to  $^{56}$Fe 
radioactive decay, while it was significantly faster beyond $\sim 320$~d.

The spectra of SN 2004et in early phase shows an H$\alpha$ profile that is
emission dominated with P-Cygni profile that is shallower than other SNe IIP.
The velocity evolution of SN 2004et determined using the weak iron lines is 
similar to other well studied SNe IIP, although the photospheric velocity in
SN 2004et was higher than other SNe IIP at all epochs.
The H$\alpha$ line indicates the presence of a high velocity component in the
early phases, with the velocity of this component being $\sim 7000$ km sec$^{-1}$ 
higher than the normal component.

The mass of $^{56}$Ni synthesized during the explosion is estimated to be
$0.06\pm 0.02\, M_\odot$.

The H$\alpha$ and [OI] 6300, 6364 \AA\ line profile evolution beyond day 320,
and the steepening of the $V$ light curve at the same epoch is interpreted as
an effect of an early dust formation.

The brightness of the plateau, its duration and the expansion velocity 
of the supernova at the middle of the plateau is used to estimate the explosion 
energy $E_{exp}=  {1.20}^{+0.38}_{-0.30} \times 10^{51}$ ergs. 
The main-sequence mass of the progenitor based on [OI] luminosity is estimated to be 
 $20\, M_\odot$. 

\section*{Acknowledgements}
We thank the anonymous referee for useful comments, which helped in improving 
the manuscript. We thank all the observers of the 2-m HCT who kindly provided part of their 
observing time for the supernova observations. We also thank Jessy Jose and 
S. Ramya for their help during observations. 
This work has made use of the NASA Astrophysics Data System and the NASA/IPAC 
Extragalactic Database (NED) which is operated by Jet Propulsion Laboratory, 
California Institute of Technology, under contract with the National 
Aeronautics and Space Administration.

\end{document}